\pdfoutput=1
\documentclass[10pt]{iopart}

\usepackage{graphicx}  
\usepackage[utf8]{inputenc}

\begin{document}

\title{White-light hyperbolic Airy beams}

\author{Andreas Valdmann, Peeter Piksarv, Heli Valtna-Lukner and Peeter Saari}

\address{University of Tartu, Institute of Physics, W. Ostwaldi Str 1, 50411, Tartu, Estonia}
\ead{andreas.valdmann@ut.ee}
\vspace{10pt}
\begin{indented}
\item[]April 2018
\end{indented}

\begin{abstract}
%Ultra-broadband Hyperbolic Airy beams are experimentally realized and spatio-temporally characterized. Transmissive (refractive) and reflective cubic phase elements were used to impose a cubic phase on the input beam. Nondispersing beams are produced in reflective geometry.
Ultra-broadband hyperbolic Airy beams are experimentally realized and spatio-temporally characterized. Transmissive (refractive) and reflective cubic phase elements were used to impose a cubic phase on the input beam. Nondispersing beams are produced in reflective geometry, while the main lobe of the hyperbolic Airy beam created with a transmissive refractive phase element suffered from lateral dispersion.

\end{abstract}

\vspace{2pc}
\noindent{\it Keywords}: Pulses, Ultrafast measurements, Propagation, Physical optics, Airy beams
%\maketitle
\ioptwocol

\section{Introduction}
The Airy beam, first realized in optics by Siviloglou \emph{et al.}, has a main intensity lobe that exhibits two surprising properties: first, it follows a curved trajectory and second, it seems to resist diffraction~\cite{Siviloglou2007b}. Since their discovery, Airy beams and their ultrashort pulsed versions~\cite{Saari2008, Abdollahpour2010, Kaganovsky2011} have been extensively studied. They have found use in a series of applications, including microscopy~\cite{Vettenburg2014, Piksarv2017}, optical micromachining~\cite{Mathis:12} and particle manipulation~\cite{Baumgartl2008}. Airy beams are typically generated by applying cubic spatial phase to a Gaussian beam by a suitable phase element and optically Fourier transforming the resulting field with a lens. The Airy beam, following a parabolic trajectory, is observed behind the lens.

However, more recently Kotlyar and Kovalev showed that if the Fourier lens is removed, a beam with the transverse intensity distribution resembling an Airy beam is formed, but the main intensity lobe of the beam follows a hyperbolic rather than parabolic trajectory~\cite{Kotlyar2014}. Unlike the regular Airy beam, the transverse scale of the so-called hyperbolic Airy (HA) beam increases during propagation.

In this paper, we present the first experimental realization and full spatio-temporal characterization of the hyperbolic Airy beam. We also show that the HA beam forms at a certain distance from the cubic phase element. Furthermore, a geometrical model is utilized to explain the temporal features of the HA pulse. 

A broadband supercontinuum light source and a SEA TADPOLE spatial-spectral interferometer are used in the experimental setup~\cite{Bowlan2006, Piksarv2012}. This combination gives us the ability to map the 3-dimensional HA light field in a single spatial scan and analyze the behaviour of different wavelength components separately similar to hyperspectral imaging. Furthermore, as the spectral phase difference induced by the cubic phase element is also recorded, the temporal evolution of the HA wave field is reconstructed and spatio-temporal effects are observed. Two different cubic phase elements---transmissive and reflective---are used to create the HA beams. It is shown that a nondispersing broadband beam is produced only in the reflective case.

\section{Hyperbolic Airy Beams and Pulses}
The HA beam is created by imposing transverse cubic phase on an incident laser beam, using, e.g., a custom phase plate or a spatial light modulator. In planar (2-dimensional) geometry the general expression for the applied phase modulation function is
\begin{equation}
\varphi_{\mathrm{2D}}(x) = \frac{x^3}{3x_0^3} + \beta x,
\label{eq:phase_1d}
\end{equation}
where $x_0$ is a lateral characteristic length of the HA field and $\beta$ is a coefficient that determines the initial deflection of the beam after the phase element.
The phase plate described above would yield a light sheet---exhibiting the properties of an HA beam only in the $x$-direction. In the experiments, however, a phase plate varying in both $x$ and $y$ directions was used to create 3-dimensional HA beams. The corresponding phase modulation function is
\begin{equation}
\varphi_{\mathrm{3D}}(x, y) = \frac{x^3+y^3}{3x_0^3} + \beta (x+y).
\label{eq:phase_2d}
\end{equation}

It has been shown in~\cite{Kotlyar2014} that if a 2-dimensional Gaussian beam with $1/e^2$ radius $w$ falls on the cubic phase element, then the resulting wave field reads
\begin{eqnarray}
&\Phi(x,z) = -\mathrm{Ai}\left[\frac{kx_0x}{z} - \beta + \left(\frac{x_0}{w}\right)^4\left(1-\frac{iz_0}{z}\right)^2\right]\nonumber \\
&\times\exp\left[\left(\frac{x_0}{w}\right)^2\left(\frac{kx_0x}{z} - \beta\right)+\frac{2}{3}\left(\frac{x_0}{w}\right)^6\left(1-3\frac{z_0^2}{z^2}\right)\right]\nonumber\\
&\times\exp\left[\frac{ikx^2}{2z}-\frac{iz_0}{z}\left(\frac{x_0}{w}\right)^2\left(\frac{kx_0x}{z} - \beta\right)\right]\nonumber\\
&\times\exp\left[\frac{2i}{3}\left(\frac{x_0}{w}\right)^6\left(3\frac{z_0}{z}-\frac{z_0^3}{z^3}\right)+ikz\right]\sqrt{\frac{-i2\pi k}{z}}x_0
\label{eq:HA_1d}
\end{eqnarray}
where $k$ is the wave number and $z_0$ is the longitudinal characteristic length. Here the assumption has been made that the beam remains paraxial after passing through the phase element. A respective 3-dimensional wave function corresponding to (\ref{eq:phase_2d}) is:
\begin{equation}
\Phi(x, y, z) = \Phi(x, z) \cdot \Phi(y, z).
\label{eq:HA_2d}
\end{equation}

So far, we have described a monochromatic HA wave field. However, it is also possible to create HA beams and pulses having a broad spectrum, as it is shown in the experimental section.
A polychromatic HA wave field is obtained by integrating over the spectrum $S(k)$:
\begin{equation}
\Psi\left(x, y, z,t\right)=\int_{0}^{\infty}\mathrm{d}kS\left(k\right)\Phi\left(x,y,z,k\right)e^{\mathrm{i}k\left(z-ct\right)}.\label{eq:HAPulse}
\end{equation}
Note that the coefficients $x_0$ and $\beta$ depend on the wavelength even in the reflective case as the phase modulation $\varphi \propto k$ when a continuous reflective surface is used. In the transmissive case, the effect of material dispersion is also added.

\begin{figure}[htbp]
  \centering
  \includegraphics{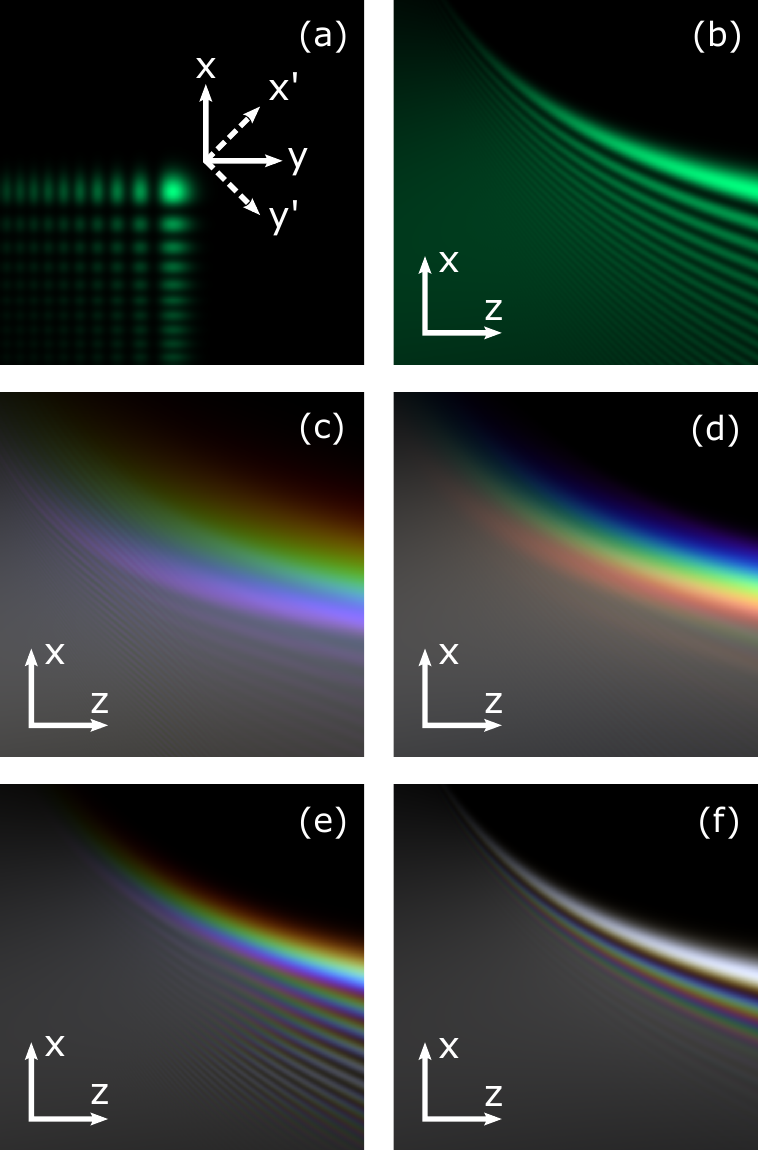}
\caption{Simulated cross section (a) and side view (b--f) of HA beams. Monochromatic (a, b) and polychromatic beams of type I, II, III, and IV are shown in (c--f), respectively. The $z$-axis is compressed by a factor of 200. We define coordinate axes $x'$ and $y'$ that are obtained by rotating the $x$ and $y$ axes around the $z$ axis by $45^\circ$ as shown in (a).}
\label{fig:Airy_types}
\end{figure}

The cross section and side view of the HA beam are shown in figure~\ref{fig:Airy_types}. A distinct pattern of interference maxima characteristic to Airy beams is apparent in the cross section of the beam. The lateral section shows the hyperbolic trajectory of the main lobe of the beam. Here, the parameter $\beta$ is set to 0 and in this case the $x$-coordinate of the main intensity maximum decreases monotonously with increasing $z$.

In the polychromatic case, the phase $\varphi$ of each wavelength component with corresponding wavenumber $k$ can be modulated with different strength, which influences the properties of the resulting beams or pulses, namely the region where the fields of all components overlap and the beam is "white". Four special types of $k$-dependence can be pointed out for HA beams~\cite{Saari2008, Kaganovsky2011}, where the spectrally overlapping region is respectively: in far field (type I, $\varphi \propto k^3$), on the cubic phase element (type II, $\varphi = \mathrm{const.}$), on the optical axis (type III, $\varphi \propto k^{3/2}$), on the trajectory of the main lobe (type IV, $\varphi \propto k$). The type IV beam would be most suitable for practical applications as its main lobe does not disperse laterally. As can be seen from figure~\ref{fig:Airy_types}, significant spatial dispersion can take place in some of the other cases. It has been previously shown, that it is possible to create type IV regular Airy beams with a refractive cubic phase plate~\cite{Valdmann2014}. We will investigate if the same holds true for HA beams.

The 3-dimensional HA beam accelerates in both $x$ and $y$ directions and the main lobe trajectory lies in the plane $x = y$. We denote $x'$ as the transverse coordinate in that plane (see figure~\ref{fig:Airy_types}(a)). 
The wavelength dependent main lobe trajectory of a 3D HA beam is~\cite{Kotlyar2014}
\begin{equation}
x' = \frac{\sqrt{2}\left(\beta(k)+y_m\right)z}{kx_0(k)} + \frac{\sqrt{2}kx_0(k)^3}{4z},
\label{eq:trajectory}
\end{equation}
where $y_m = -1.01879$ is the first maximum of the Airy function. In the reflective case $\beta \propto k^{2/3}$ and $x_0 \propto k^{-1/3}$. In the refractive case $\beta$ and $x_0$ also depend on the material dispersion of the cubic phase element.

\section{Finite Aperture HA Beams}
Next we show that the main lobe of a finite energy HA beam is not observable directly behind the cubic phase element and a transition region exists, where the initial intensity distribution of the incoming beam is transformed into the distinct Airy pattern.%, where the main lobe is distinguished from the rest of the beam.

\begin{figure}[htbp]
  \centering
  \includegraphics{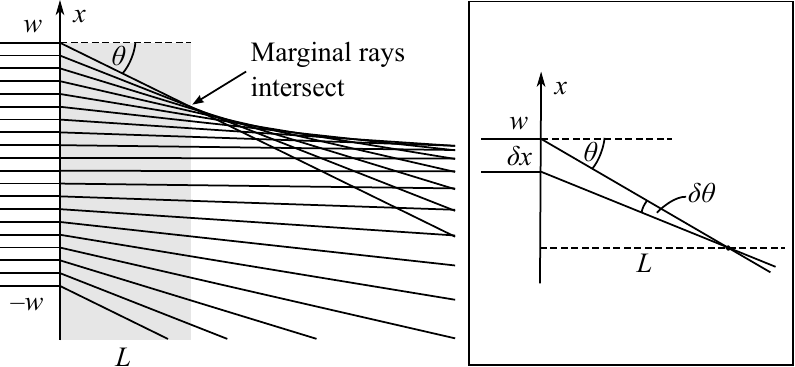}
\caption{Formation of the HA beam. Left: a collimated beam passes through a thin cubic phase mask on the x-axis. Main lobe of the hyperbolic Airy beam is formed after the transition region (gray) with length $L$. Right: calculating $L$ using marginal rays. Two rays originally separated by small distance $\delta x$ form a small angle $\delta\theta$ after passing through the phase mask and intersect at distance $L$ from the phase element.}
\label{fig:hyperbolic_Airy_caustics}
\end{figure}

A geometrical model is used to estimate the length $L$ of the transition region. Consider a collimated light beam with a finite width $2w$ falling on the centre of a thin 1D cubic phase mask (see figure~\ref{fig:hyperbolic_Airy_caustics}). At the longitudinal distance $L$ from the phase mask, the rays originating from the edge of the beam begin to intersect and a caustic is formed which corresponds to the main intensity lobe of the HA beam. To find $L$, we look at two rays whose initial coordinates on the phase mask are $x_{02} = w$ and $x_{01} = w-\delta x$. The rays are deflected by angles $\theta_2$ and $\theta_1$ that depend on the phase modulation function $\varphi(x)$ as
\begin{equation}
\theta(x_0) = \arcsin\left(k^{-1}\frac{d\varphi}{dx}\bigg\vert_{x=x_0}\right).
\label{eq:theta}
\end{equation}
The angle between the rays after passing through the phase mask is $\delta \theta = \theta_2 - \theta_1$. In the paraxial case, where $\theta \ll 1$, we can use the approximation $\delta x \approx L\delta\theta$. In that case (\ref{eq:theta}) also simplifies as $\sin(\theta) \approx \theta$. When the rays are close, i.e., $\delta x \rightarrow 0$, then $1/L = d\theta / dx$ and
\begin{equation}
L = k\left[\frac{d^2\varphi}{dx^2}\bigg\vert_{x=w}\right]^{-1}.
\label{eq:L_general}
\end{equation}
This is a general expression for the length of the transition region in the 2D paraxial case for any monotonous phase modulation function $\varphi(x)$. For a cubic modulation function given by (\ref{eq:phase_1d}), we obtain
\begin{equation}
L_{2\mathrm{D}} = \frac{3}{2}kx_0^3w^{-1}.
\end{equation}
If the cubic function is 2-dimensional, then the derivative in (\ref{eq:L_general}) is replaced by the gradient of a 2D phase modulation function evaluated at the point $(w/\sqrt{2},\, w/\sqrt{2})$. The resulting characteristic distance is
\begin{equation}
L_{3\mathrm{D}} = \frac{3\sqrt{2}}{2}kx_0^3w^{-1}.
\end{equation}

%\begin{figure}[htbp]
%  \centering
%  \includegraphics{"fig_L_derivation_paraxial"}
%\caption{Two rays originally separated by small distance $\delta x$ pass through a thin cubic phase element on the x-axis. The rays intersect at distance $L$ from the phase element.}
%\label{fig:L_derivation_paraxial}
%\end{figure}

\section{Propagation of the HA pulse front}

When analysing the spatio-temporal behaviour of ultrashort laser pulses with seemingly complex spatial structure, it is sometimes useful to look at the evolution of the pulse directly in the spatio-temporal domain, rather than trying to put together the intricate interference patterns of its monochromatic constituents. Simple geometrical considerations can often be used to explain the key features of the pulse. For example, it has been shown how the Arago spot is formed in the temporal domain as a toroidal boundary wave intersects with itself after an ultrashort laser pulse falls on a circular disk~\cite{Piksarv2011}.

\begin{figure}[htbp]
  \centering
  \includegraphics[width=0.6\linewidth]{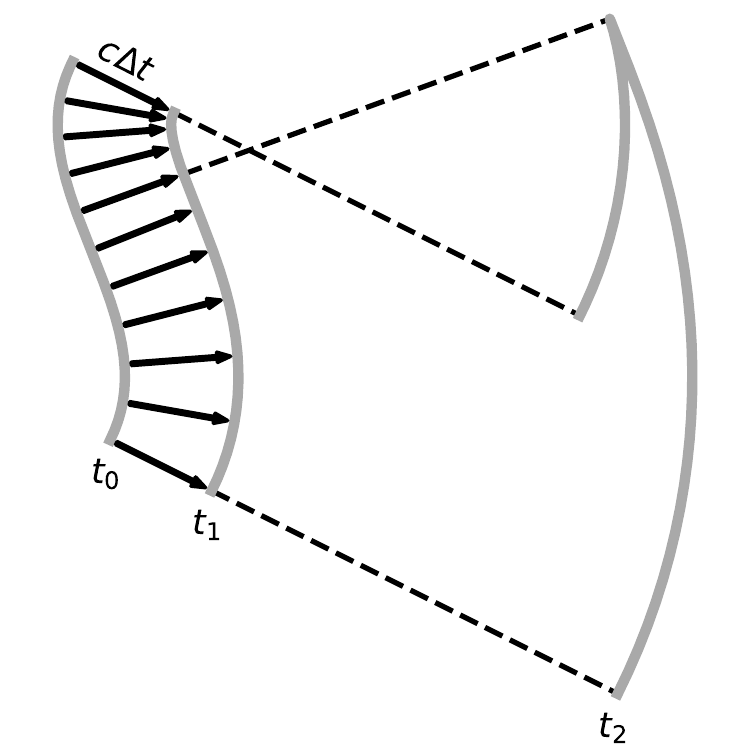}
\caption{Propagation of a hyperbolic Airy pulse front in the 2D case. The pulse front (gray) has a cubic profile at time $t_0$. Lines normal to it with length $c\Delta t$ are drawn to construct the pulse front at $t_1$. By time $t_2$ the pulse front has folded on itself and a caustic is formed at the fold.}
\label{fig:pulse_fronts_2D}
\end{figure}

Here we use an intuitive geometrical model to explain the temporal structure of the side lobe tails that accompany the main lobe of an ultrashort HA pulse. Imagine a short plane wave pulse falling on a mirror with cubic surface profile in the x-direction (at first we look at the 2D case and assume that the shape of the mirror is not changing in the y-direction). Right after reflection, the pulse front has attained a cubic shape with twice the displacement of the mirror profile relative to a plane wave. Let us call this the initial pulse front at time $t_0$ (see figure~\ref{fig:pulse_fronts_2D}).

Next, we use a ray-based method to find the shape of the pulse front at time $t_0 + \Delta t$. The rays originating from the initial pulse front are pointed along the local phase gradient, i.e., they are normal to the pulse front. We draw rays with length $c\Delta t$ and the resulting pulse front is obtained by connecting the end points of the rays with a smooth curve. It has been shown in~\cite{Kaganovsky2010, Kaganovsky2011} that this model can be extended with the theory of uniform geometrical optics to yield the exact spatio-temporal form of the pulse front in a rigorous calculation. However, here we concentrate on the simpler model to retain the intuitive aspect of the explanation.

When $c\Delta t$ is small compared to the local curvature radius of the pulse front, then the resulting pulse front is only slightly distorted from the original and retains its overall smooth shape. This corresponds to the transition region right after the cubic phase element discussed in the previous section.  However, when $c\Delta t$ exceeds the minimum curvature radius of initial pulse front, the pulse front starts to fold on itself and a caustic is formed. After some propagation, two distinct tails are developed in the side lobe region. This results in a double pulse in the temporal domain when the pulse tail is sampled with a small aperture (e.g., a pinhole or a fibre tip). It should be noted that the temporal double tail of the pulse persists when a lens is used to transform the HA pulse into a regular Airy pulse. This feature of the Airy pulse has been theoretically predicted in~\cite{Saari2008} and~\cite{Kaganovsky2011}.

\begin{figure}[htbp]
  \centering
  \includegraphics[width=0.9\linewidth]{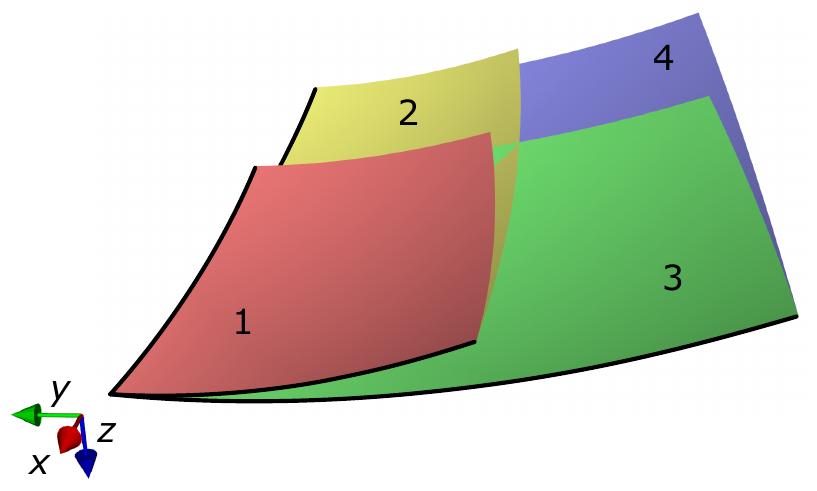}
\caption{3D hyperbolic Airy pulse front after propagation outside of the transition region. The pulse front is folded into four sheets with sheets 2 and 3 intersecting. Caustics are formed at the folds (black lines). Main lobe of the pulse is located near the corner with the coordinate triad.}
\label{fig:pulse_front_3D}
\end{figure}

We used the same model to examine the pulse front propagation of the HA pulse in the full 3D case. The initial cubic pulse front described by (\ref{eq:phase_2d}) was given on a square grid in $x$ and $y$ directions. The resulting HA pulse front after propagation is shown in figure~\ref{fig:pulse_fronts_2D}. Here a total of four sheets are formed originating from different quadrants of the initial pulse front. Notice the sheets are connected on the edges as they are parts of the same pulse front that has been folded over along the caustic lines---similar to folding paper in Origami. However, unlike physical paper, the two centre sheets intersect and form a cross pattern when the pulse is cut with a plane that is perpendicular to the $x’$-axis.

Finally, it should be noted that the geometrical model works well in situations where the monochromatic constituents of the beam have coinciding wave fronts. This is the case, for example, when a short plane wave pulse reflects from a mirror surface or diffracts from an edge. However, if the wave vectors of the monochromatic constituents  start to diverge (e.g., after diffracting from a grating), the initially sharp pulse front will spread out in time and other methods should be used to characterize the propagation of the pulse front. 

\section{Experimental setup}
Two different optical components were used to impose the cubic phase on the beam. First, a moulded PMMA free-form element with a continuous surface profile was used in transmissive geometry. To ensure that the HA beam is formed reasonably close to the phase element, and that its hyperbolic trajectory is observable, a significantly steeper phase profile was chosen, than has been previously used for generating regular Airy beams, e.g., in~\cite{Valdmann2014}. The second (reflective) phase element was obtained by coating the cubic surface of an identical element with silver. The refractive index of PMMA is close to 1.5 in the measurement wavelength range and therefore, the phase modulation depth is effectively four times higher in the reflective geometry.

\begin{figure}[htbp]
  \centering
  \includegraphics{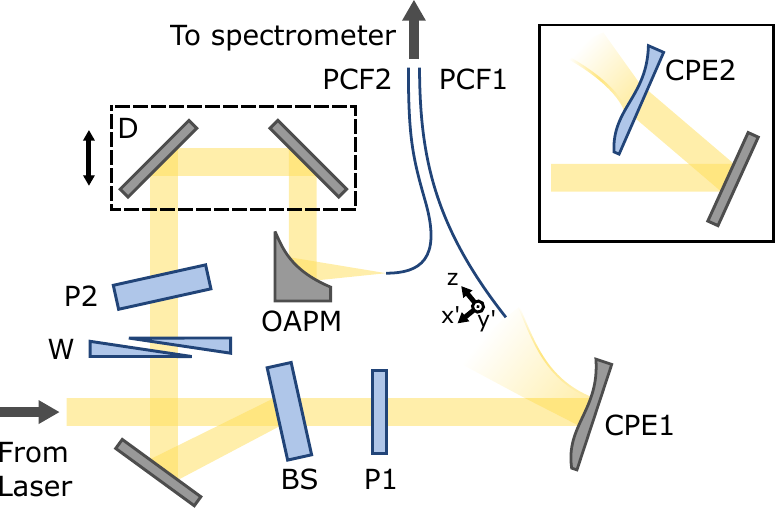}
\caption{Experimental setup with reflective (main figure) and transmissive (insert) cubic phase element. The beam is split into two arms of the interferometer with a fused silica window (BS) used as a beam splitter. The dispersion in both arms is balanced with fused silica windows (P1, P2) and wedges (W). The cubic phase element (CPE) is placed into the measurement arm and the produced HA beam is scanned with a PCF fibre tip. In the reference arm a variable delay line (D) is used and the beam is focused onto the PCF tip with an off-axis parabolic mirror (OAPM). The output ends of PCF1 and PCF2 are fed into a custom built spectrometer that records the spectral interference pattern.}
\label{fig:setup}
\end{figure}

The experimental setup is shown in figure~\ref{fig:setup}. We used a SEA TADPOLE \cite{Bowlan2006} type spatial-spectral interferometer to obtain the spatio-temporal impulse response of the cubic phase elements. Fianium supecontinuum laser source (SC400-2-PP) was used to cover an ultra-wide spectral range of 450--950 nm in a single measurement, which results in few-femtosecond temporal resolution. The supercontinuum laser beam was spatially filtered and expanded using a length of photonic crystal fibre (PCF) and off-axis parabolic mirrors.

The cubic phase element was placed into the measurement arm of the spectral interferometer and the field behind it was scanned with a photonic crystal fibre tip. The same setup was used to yield the main lobe trajectories of the different wavelength components.

\section{Simulation methods}
The simulations were carried out by numerically evaluating (\ref{eq:HA_1d}) to (\ref{eq:HAPulse}). The input parameters were obtained as follows. The beam waist size was measured with a CCD camera at 7 wavelengths ranging from 450-1000 nm and a linear fit was made between the wavelength and waist size. The surface profile of the cubic phase element was measured with a high accuracy CNC coordinate measuring machine and the lateral characteristic length $x_0(k)$ was derived from the cubic coefficient of the measured profile. The longitudinal characteristic length $z_0$ is given as $z_0(k) = kw^2/2$. The parameter $\beta(k)$ was obtained by fitting the experimental beam trajectory to (\ref{eq:trajectory}). In the transmissive case, the dispersion curve of PMMA and the linear component of the measured surface profile were taken into account when finding $x_0(k)$ and $\beta(k)$. Spectral amplitude $S(k)$ was chosen according to the experimentally measured spectrum.

\section{Results and Conclusion}

%\begin{figure*}[tbh]
%  \begin{centering}
%  \includegraphics{fig_2}
%  \par\end{centering}
%  \caption{Measured intensity cross sections of white-light HA beams. Transmissive (top) and reflective (bottom) free-form elements were used to impose the cubic phase on the beam. Intensity cross-section of full spectrum (white) beam  (a, e) in pseudocolour and of its selected spectral components: 500 nm (b, f), 600 nm (c, g), and 700 nm (d, h).}
%  \label{fig:cross_section}
%\end{figure*}

%\begin{figure}[htbp]
%  \begin{centering}
%  \includegraphics[width=\linewidth]{fig_cross_section_rgb_ab}
%  \par\end{centering}
%  \caption{Measured intensity cross sections of white-light HA beams. Transmissive (a) and reflective (b) free-form elements were used to impose the cubic phase on the beam.}
%  \label{fig:cross_section}
%\end{figure}

\begin{figure*}[tbh]
  \begin{centering}
  \includegraphics[width=0.9\linewidth]{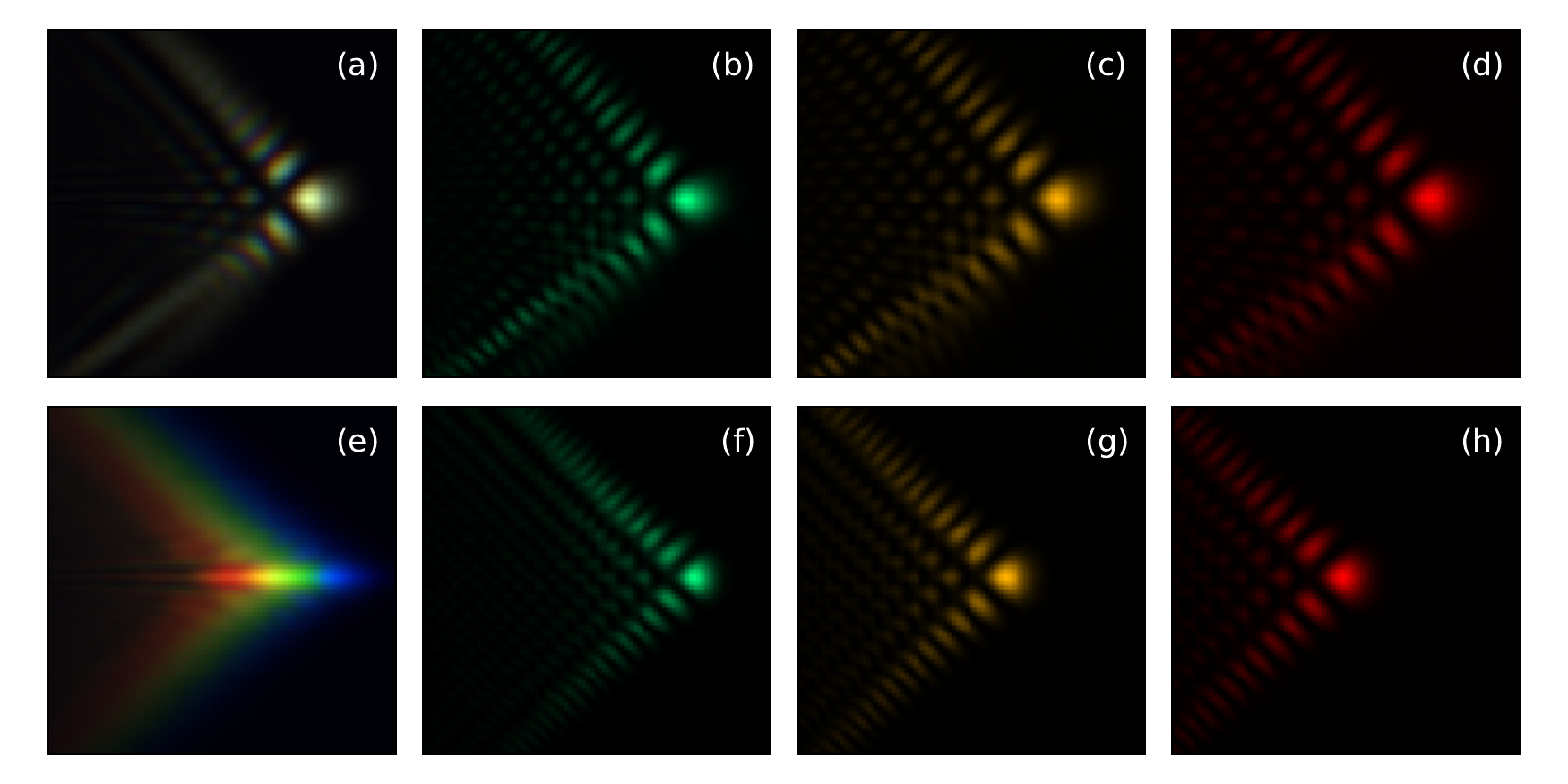}
  \par\end{centering}
  \caption{Measured hyperspectral intensity cross sections of white-light HA beams. Transmissive (top) and reflective (bottom) free-form elements were used to impose the cubic phase on the beam. Composite image with all the spectral components is shown (a, e) along with  selected wavelengths: 500 nm (b, f), 600 nm (c, g), and 700 nm (d, h).}
  \label{fig:cross_section}
\end{figure*}

The measured intensity cross sections of HA beams, created with transmissive and reflective cubic phase elements, are shown in figure~\ref{fig:cross_section}. In the transmissive case, the main lobe of the beam is smeared out and laterally dispersed. It becomes apparent that material dispersion of the cubic phase element has to be taken into consideration, when its thickness profile varies significantly to fulfill the preconditions for creating HA beams.

However, when a reflective element was used, the main lobe of the beam did not disperse, i.e., a type IV HA beam was produced. For type IV HA beams, the spacing of the side lobes depends on the wavelength, and therefore they become less defined as the interference maxima and minima of different wavelength components start to overlap. In fact, this could be favourable in some applications as the relative intensity of the main lobe increases in respect to the side lobe background.

\begin{figure}[htbp]
  \centering
  \includegraphics[width=0.95\linewidth]{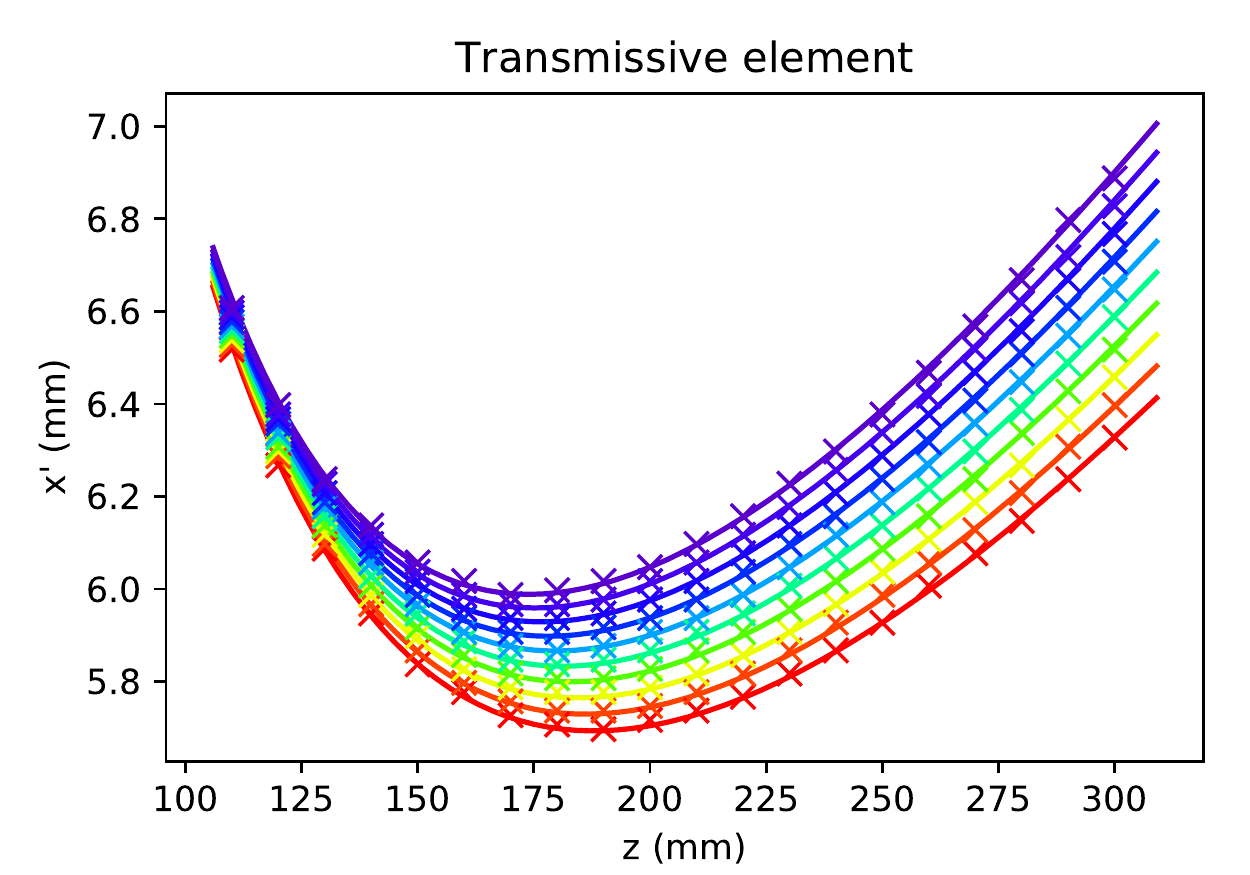}
  \caption{Hyperbolic Airy beam trajectory with a transmissive cubic phase element. Theoretical (solid lines) and experimental (crosses) trajectories are shown for discrete wavelengths in range 490--900 nm. The main lobe is spread out and spatially chirped due to material dispersion.}
  \label{fig:trajectory_refr}
\end{figure}

\begin{figure}[htbp]
  \centering
  \includegraphics[width=0.95\linewidth]{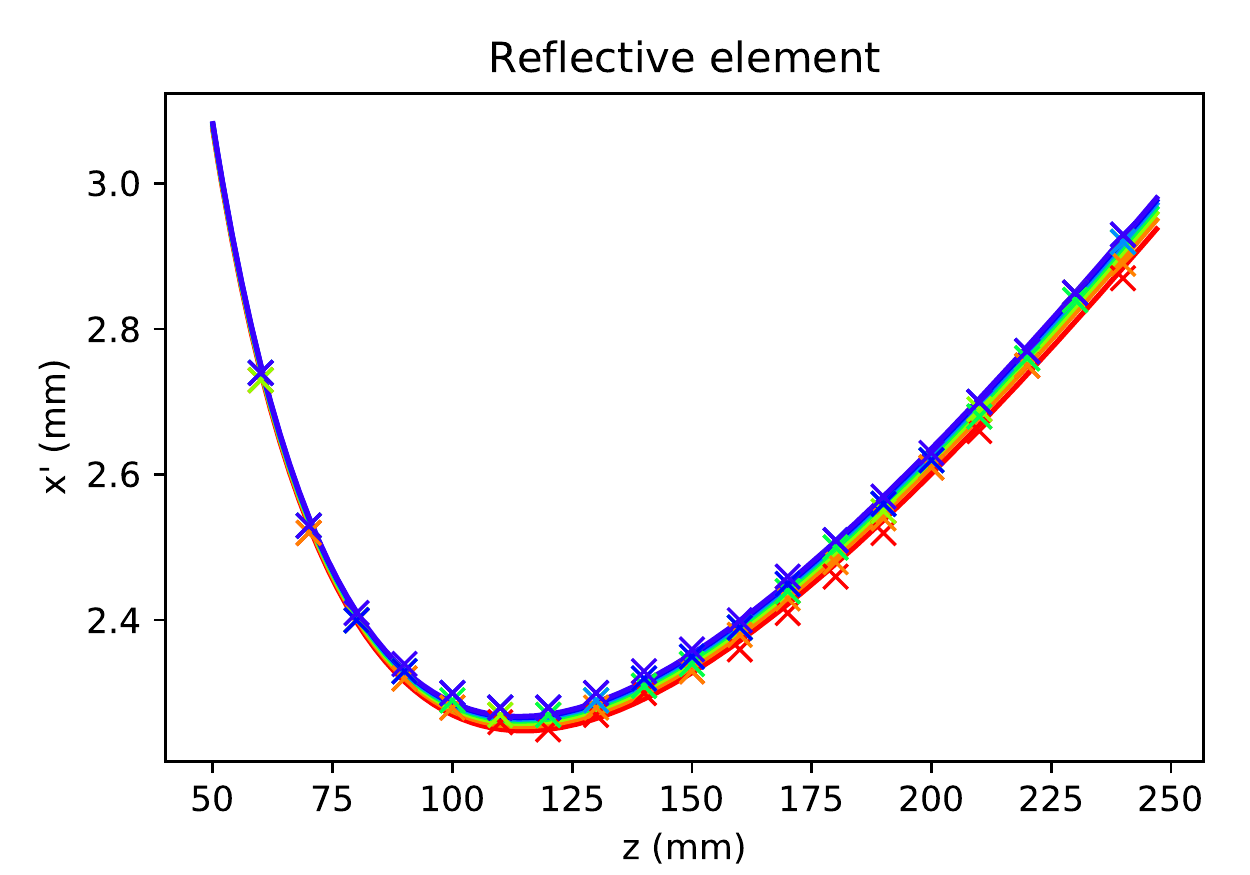}
  \caption{Hyperbolic Airy beam trajectory with a reflective cubic phase element. Theoretical (solid lines) and experimental (crosses) trajectories are shown for discrete wavelengths in range 490--900 nm.}
  \label{fig:trajectory_refl}
\end{figure}

The experimental and theoretical beam trajectories for different wavelengths are compared in figures \ref{fig:trajectory_refr} and \ref{fig:trajectory_refl}. As in the cross section views, it is seen that the main lobe of the beam is suffering from lateral dispersion  in the transmissive setup. In the reflective case the spectral chirp is greatly reduced, although a slight spread still remains as the perfect overlap is achieved at the point where the argument of the Airy function is 0. The distance of intensity maximum from this point scales with the wavelength and therefore the main lobe trajectories of different wavelengths diverge slightly.

\begin{figure*}[p]
  \centering
  \includegraphics[width=\linewidth]{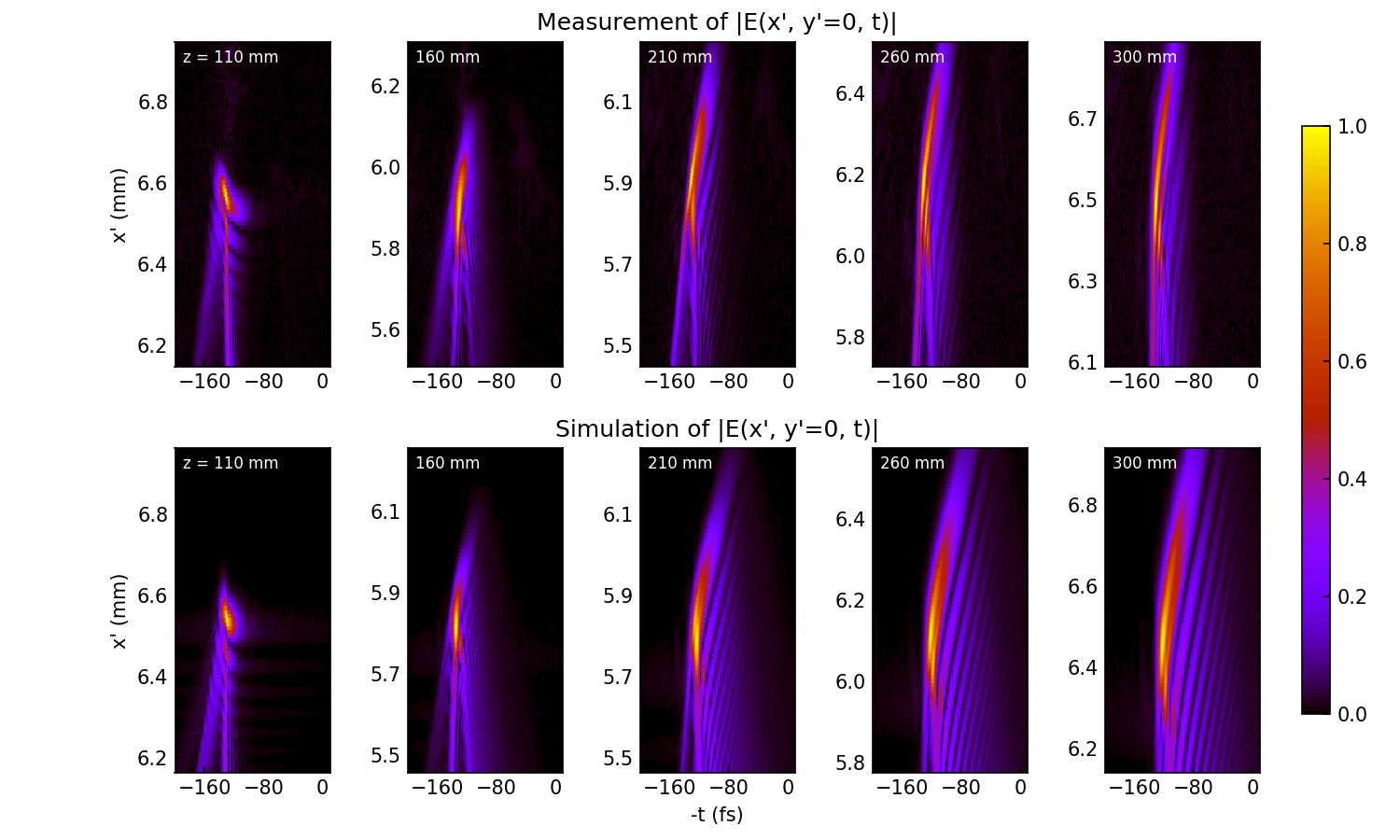}
  \caption{Measured (top) and simulated (bottom) HA field with the transmissive cubic phase element. The magnitude of the electric field is shown at different propagation distances. The reference frame is laterally comoving with the intensity maximum of the field.}
\label{fig:et_refr}
\end{figure*}

\begin{figure*}[p]
  \centering
  \includegraphics[width=\linewidth]{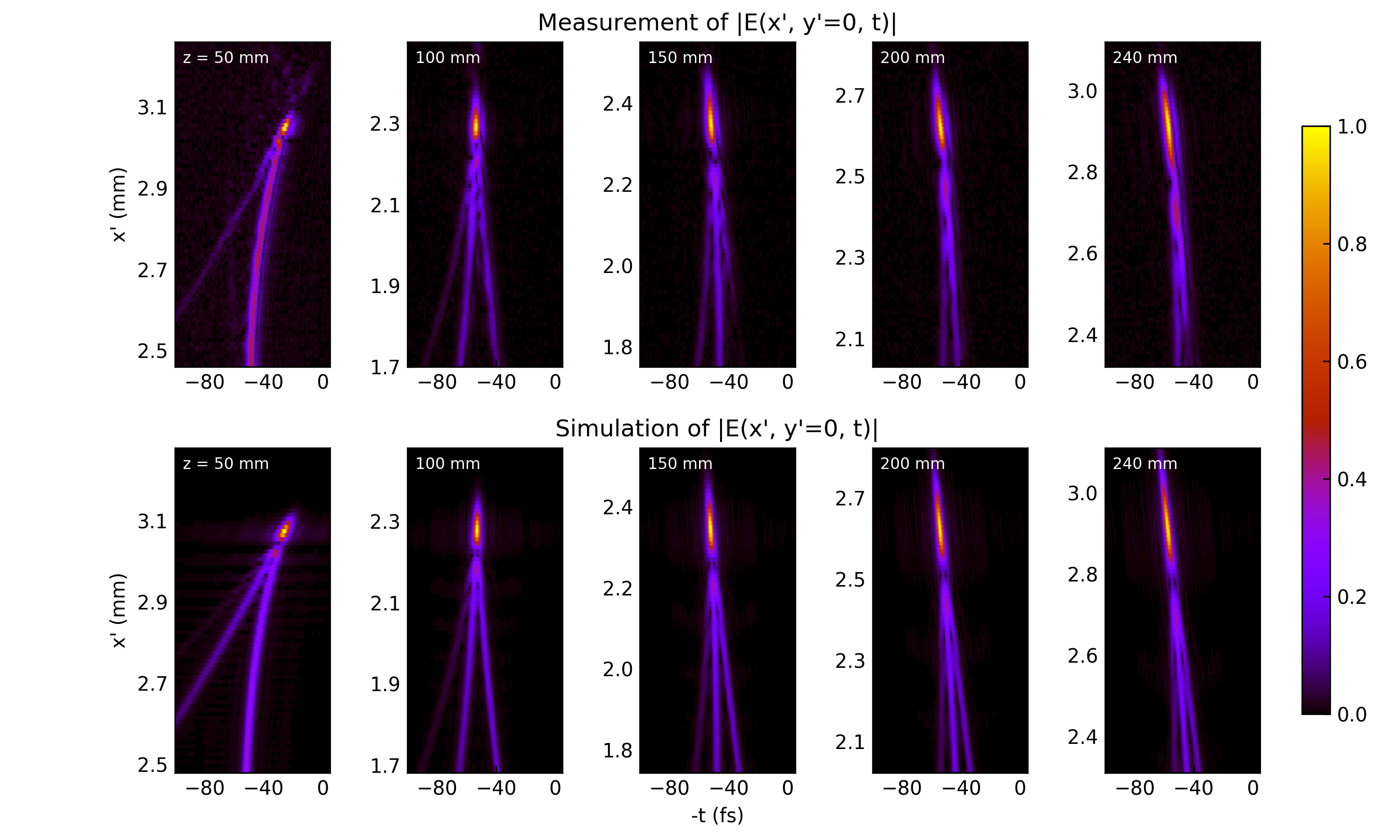}
  \caption{Measured (top) and simulated (bottom) HA field with the reflective cubic phase element. The magnitude of the electric field is shown at different propagation distances. The reference frame is laterally comoving with the intensity maximum of the field.}
\label{fig:et_refl}
\end{figure*}

\begin{figure}[tbh]
  \centering
  \includegraphics[width=\linewidth]{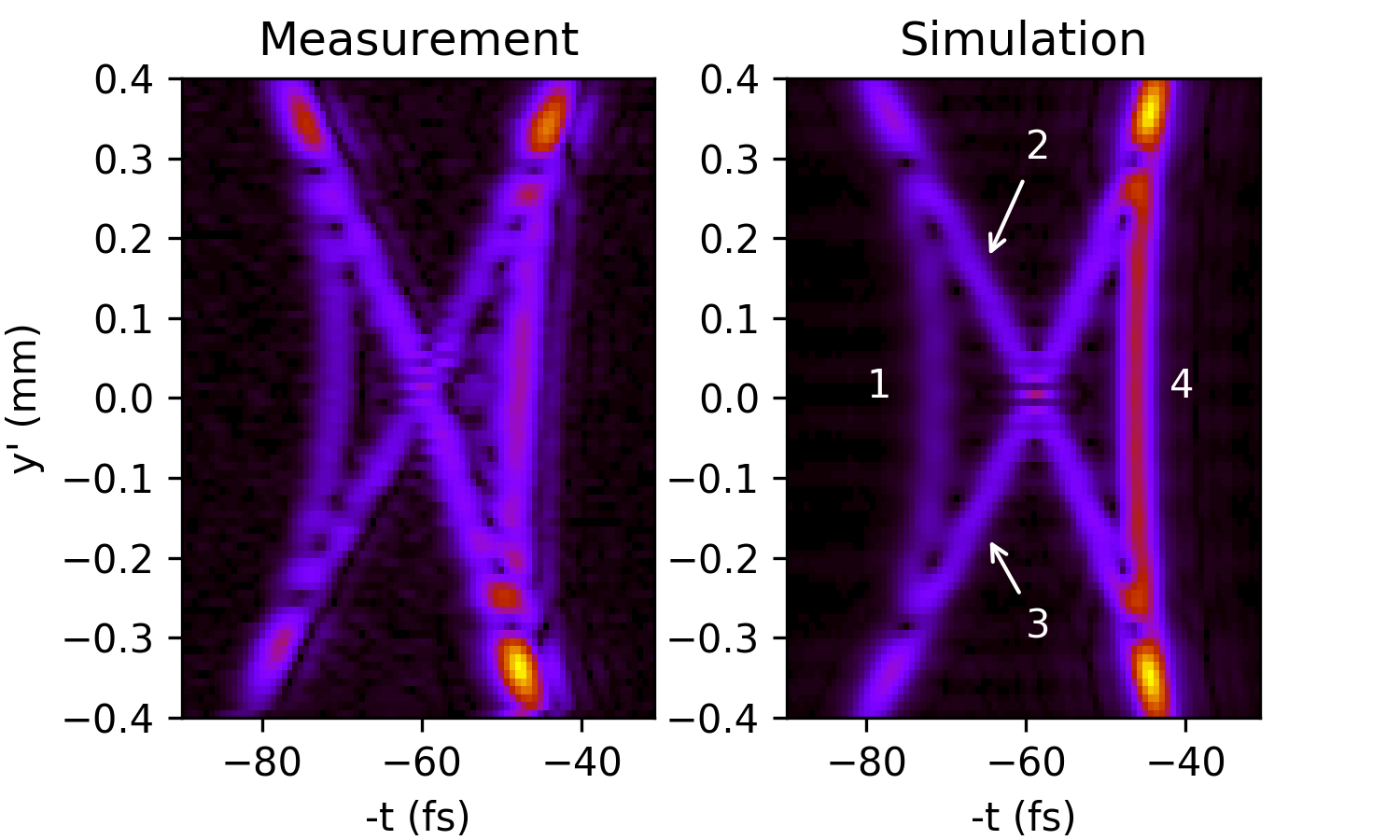}
  \caption{Measured (left) and simulated (right) section of the HA field at $x'=1.93$~mm and $z=100$~mm, created with a reflective element. Four sheets of the pulse front are numbered according to figure~\ref{fig:pulse_front_3D}.}
  \label{fig:y_section}
\end{figure}

The temporal evolution of the HA field is shown in figures~\ref{fig:et_refr} and \ref{fig:et_refl}. The complete measured data set is also shown as animations in Media 1 and Media 2.  Due to spectral narrowing, the temporal duration of the main intensity lobe was stretched from 3 fs to 17 fs with the transmissive element. With reflective geometry a distinct 3-tailed side lobe structure of the pulse front is visible in the temporal plots. This can be explained when comparing with figure~\ref{fig:pulse_front_3D}. The outer tails correspond to pulse front sheets 1 and 4, while the center tail is formed along the line where sheets 2 and 3 intersect. To further illustrate the result, a lateral section along $y'$ is shown in figure~\ref{fig:y_section}. Here all 4 sheets of the pulse front are distinguished.

In conclusion, we have experimentally realized white-light hyperbolic Airy beams created with transmissive (refractive) and reflective cubic phase elements. Using a refractive phase element results in significant lateral dispersion of the main intensity lobe, which in turn causes temporal broadening. We showed that these issues are solved by using a reflective phase element, in which case the main intensity lobe of the hyperbolic Airy beam does not disperse even when an ultra-broadband light source is used. We have also shown that in case of finite energy HA beams, the main lobe of the beam becomes observable only after a certain distance from the cubic phase element. This distance is inversely proportional to the second derivative of the phase function imposed on the input beam. Finally, an intuitive geometrical model was used to explain the layered temporal structure of the pulse front observed in the experiments.

\section*{Acknowledgements}
We thank David Crosby from Eyejusters Ltd and Viktorija Pasko from Adlens Ltd for providing the cubic phase elements. This work has been supported by the Estonian Research Council (PUT369, PUT1075).

\section*{References}

%\bibliographystyle{unsrt}
%\bibliography{Valdmann_OL_2018}

%\begin{thebibliography}{10}
%\newcommand{\enquote}[1]{``#1''}
%\end{thebibliography}

\end{document}